\documentclass[aps,prl,twocolumn,amsmath,amssymb,floatfix,superscriptaddress]{revtex4-1}
\usepackage{tabularx}
\usepackage{graphicx}
\usepackage{euscript}
\usepackage{epsfig,psfrag,subfigure}
\usepackage{dcolumn}
\usepackage{bm}
\usepackage{setspace}
\usepackage{floatrow}
\usepackage{appendix}
\usepackage{color}
\usepackage{amsfonts}
\usepackage{exscale}
\usepackage{multirow}
\usepackage{hyperref}
\hypersetup{colorlinks=false,linkcolor=black}
\usepackage[left]{lineno}
\usepackage{comment}
\usepackage{ulem}
\usepackage[dvipsnames]{xcolor}
\def\be{\begin{equation}}       \def\ee{\end{equation}}
\def\bea{\begin{eqnarray}}      \def\eea{\end{eqnarray}}

\newcommand{\PreserveBackslash}[1]{\let\temp=\\#1\let\\=\temp}
\newcolumntype{C}[1]{>{\PreserveBackslash\centering}p{#1}}
\usepackage{ulem}
\begin{document}
\captionsetup[figure]{labelfont={bf}}

\newcommand\redsout{\bgroup\markoverwith{\textcolor{red}{\rule[0.5ex]{2pt}{0.4pt}}}\ULon}
\newcommand{\red}[1]{\textcolor{red}{#1}}

\title{Charge order and superconductivity in a minimal two-band model for infinite-layer nickelates}

\author{Cheng Peng}
\affiliation{Stanford Institute for Materials and Energy Sciences (SIMES), SLAC National Accelerator Laboratory, Menlo Park, CA 94025, USA}
\author{Hong-Chen Jiang}
\email{hcjiang@stanford.edu}
\affiliation{Stanford Institute for Materials and Energy Sciences (SIMES), SLAC National Accelerator Laboratory, Menlo Park, CA 94025, USA}
\author{Brian Moritz}
\affiliation{Stanford Institute for Materials and Energy Sciences (SIMES), SLAC National Accelerator Laboratory, Menlo Park, CA 94025, USA}
\author{Thomas P. Devereaux}
\affiliation{Stanford Institute for Materials and Energy Sciences (SIMES),
 SLAC National Accelerator Laboratory, Menlo Park, CA 94025, USA}
\affiliation{Department of Materials Science and Engineering, Stanford University, Stanford, California 94305, USA}
\author{Chunjing Jia}
\email{chunjing@stanford.edu}
\affiliation{Stanford Institute for Materials and Energy Sciences (SIMES),
  SLAC National Accelerator Laboratory, Menlo Park, CA 94025, USA}

\begin{abstract}
The recent discovery of superconductivity in infinite-layer nickelates has drawn considerable attention; however, a consensus on the fundamental building blocks and common ingredients necessary to understand and describe their ground states and emergent properties is lacking. A series of experimental and theoretical studies have suggested that an effective two-band Hubbard model with Ni 3$d_{x^2-y^2}$ and rare-earth ($R$) 5$d$ character may describe the low-energy physics. Here, we study the ground state properties of this two-band model on four-leg cylinders using the density-matrix renormalization group (DMRG) technique to better grasp whether such a simple model can embody the essential physics. A key difference compared to single-band Hubbard materials is that the system is self-doped: even at overall half-filling, the $R$-band acts as an electron reservoir, hole-doping the Ni-layer,  and fundamentally altering the physics expected from an undoped antiferromagnet. On the four-leg cylinder, the ground state is consistent with a Luttinger liquid, with anti-phase modulations of the charge density in the Ni- and $R$-layers having corresponding wavevectors that lock together. Light hole doping away from 1/2 filling releases the locking between the Ni and the $R$ charge modulations, as the electron density in the $R$-band decreases and eventually becomes exhausted at a hole doping concentration that depends sensitively on the effective splitting between the Ni and the $R$ orbitals. The ground state of the doped system is consistent with a Luther-Emery liquid, possessing quasi-long-range superconducting correlations in the Ni layer, similar to the single-band Hubbard model. Our results are consistent with experimental observations and may help to reveal the microscopic mechanism for pairing and other emergent properties not only in the infinite-layer nickelates but also other unconventional superconductors.
\end{abstract}
\maketitle

\section{Introduction}
Long sought as close relatives of the cuprates, superconductivity with a critical temperature $T_c\sim 10$\,K was discovered recently in a series of infinite-layer nickelate materials synthesized by topotactic reduction (R$_{1-x}$Sr$_{x}$NiO$_2$, where $R$ = La, Nd, Pr)~\cite{Li2019,Osada2020, Zeng2020,Li2020, Osada2021,2022arXiv220302580L}. Unlike the cuprates, measurements of the Hall coefficient have shown that the charge carriers change from electron-like to hole-like upon doping, suggesting the presence of an electron pocket in the parent compounds~\cite{Li2020}. At the same time, both resonant inelastic x-ray scattering (RIXS) and electron energy loss spectroscopy (EELS) measurements have highlighted the essential contributions from the rare-earth 5$d$ orbitals~\cite{Rossi2020, Goodgee2021pnas}. These rare-earth itinerant electrons, which might interact or hybridize strongly with the more correlated magnetic Ni orbitals, already may lead to new physics absent in the cuprates~\cite{Hepting2020}. For example, very recent resonant X-ray scattering experiments have revealed the presence of charge order in the undoped  Nd- and La-based parent compounds~\cite{Matteo2021arxiv,ZhouKeJin2021arxiv,Krieger2021arxiv,2022arXiv220302580L}, where one normally would find strong antiferromagnetic order as in the undoped cuprate family. Even more unexpectedly, there also are indications of a superconducting state in undoped LaNiO$_2$ without rare-earth magnetism\cite{Osada2021,2022arXiv220302580L}. These discoveries, especially in the undoped parent compounds, truly distinguish the infinite-layer nickelates from their close relatives, the cuprates.

The ingredients of an effective low-energy model can be further constrained by additional experimental results. For example, X-ray absorption spectroscopy (XAS) experiments near the Ni $L$ edge absorption have shown that the Ni $3d_{x^2-y^2}$ orbital plays a dominant role upon doping~\cite{Rossi2020}, as well as suggesting that the Ni $3d_{z^2}$ orbital plays only a minor role. Other XAS measurements at the oxygen $K$ edge have not observed the characteristic signatures of oxygen density of states near the Fermi energy~\cite{Rossi2020} with hole doping, likely indicating that these nickelates are in the Mott-Hubbard limit, rather than the charge-transfer limit, of the Zaanen-Sawatzky-Allen (ZSA) scheme~\cite{Zaanen1985prl}. On the other hand, EELS measurements~\cite{Goodgee2021pnas} have demonstrated an emergent hybridization, reminiscent of a Zhang–Rice singlet (ZRS) and oxygen-projected states; however, the spectral weight of the O-$2p$ features remains rather small, even at high doping levels. Such experimental findings provide additional evidence that these infinite-layer nickelates are distinct from the cuprates, where ZRSs dominate the low-energy physics~\cite{Chen1991prl} and much of the theoretical work has been focused on effective single- and multi-orbital Hubbard models of the physics restricted to the two-dimensional CuO$_2$-planes.

While debate remains about the most appropriate minimal effective model, 
the electronic structure of the parent infinite-layer nickelates has been discussed for quite some time~\cite{Anisimov1999,Lee2004}. Soon after the discovery of superconductivity, various theoretical models were proposed to describe the low-energy physics.~\cite{Hirayama2020,Botana2020prx,Wu2020prb,Meijw2020,Sakakibara2020prl,Lechermann2020,WuX2020,Nomura2020,Jiang2020,Hepting2020,Sakakibara2020prl,Wang2020prb,Werner2020prb,Bhattacharyya2020prb,Liu2021prb,Bjornson2021PRB,Zhang2020prr,Fu2019arxiv,Nomura2019prb}. Whether or not Kondo physics is relevant in these materials remains unclear; however, the experimental results provide strong evidence that multi-orbital models, which include a $d_{x^2-y^2}$ orbital in the Ni-O plane and a $R$ 5$d$ orbital of itinerant electrons, could provide an effective electronic description of the parent state of these materials. Consistency between model predictions and experimental observations can further test the validity of such effective Hamiltonians.

Here, we use the numerically unbiased density matrix renormalization group (DMRG)\cite{White1992} method to study the ground state properties of a two-orbital model and examine its relevance to infinite-layer nickelates. Specifically, we focus on the effect of the  rare-earth 5$d$ orbital and Ni 3$d_{x^2-y^2}$ orbital in both the half-filled parent and doped cases. In the parent compound (or at half-filling to be consistent with the convention in cuprates), we find that the ground state of quasi-one-dimensional four-leg cylinders is consistent with a Luttinger liquid (LL), characterized by quasi-long-range charge and spin correlations in the $R$-layer but short-range correlations in the Ni layer. Electrons in the $R$-layer, responsible for the electron pocket, hybridize with the correlated Ni orbitals, self-doping holes into the Ni layer. This gives rise to a charge density wave (CDW) in both layers at half-filling, with an anti-phase modulation of the charge density between the Ni- and $R$-layers and corresponding wavevectors proportional to the carrier concentration that lock together. With hole doping, the $R$-layer empties and becomes insulating (at a hole doping concentration $\sim 12.5\%$ for the parameters shown here); and the ground state of the system is consistent with a Luther-Emery(LE) liquid \cite{LE1974}, with quasi-long-range superconducting (SC) and CDW correlations in the Ni-layer but short-range spin and single-particle correlations. Consistency between our results and experiments provides additional support for the relevance of this two-band effective model to the low-energy physics in the infinite-layer nickelates.

\section{Model and Method}

\begin{figure*}[tp]
\centering
    \includegraphics[width=1\linewidth]{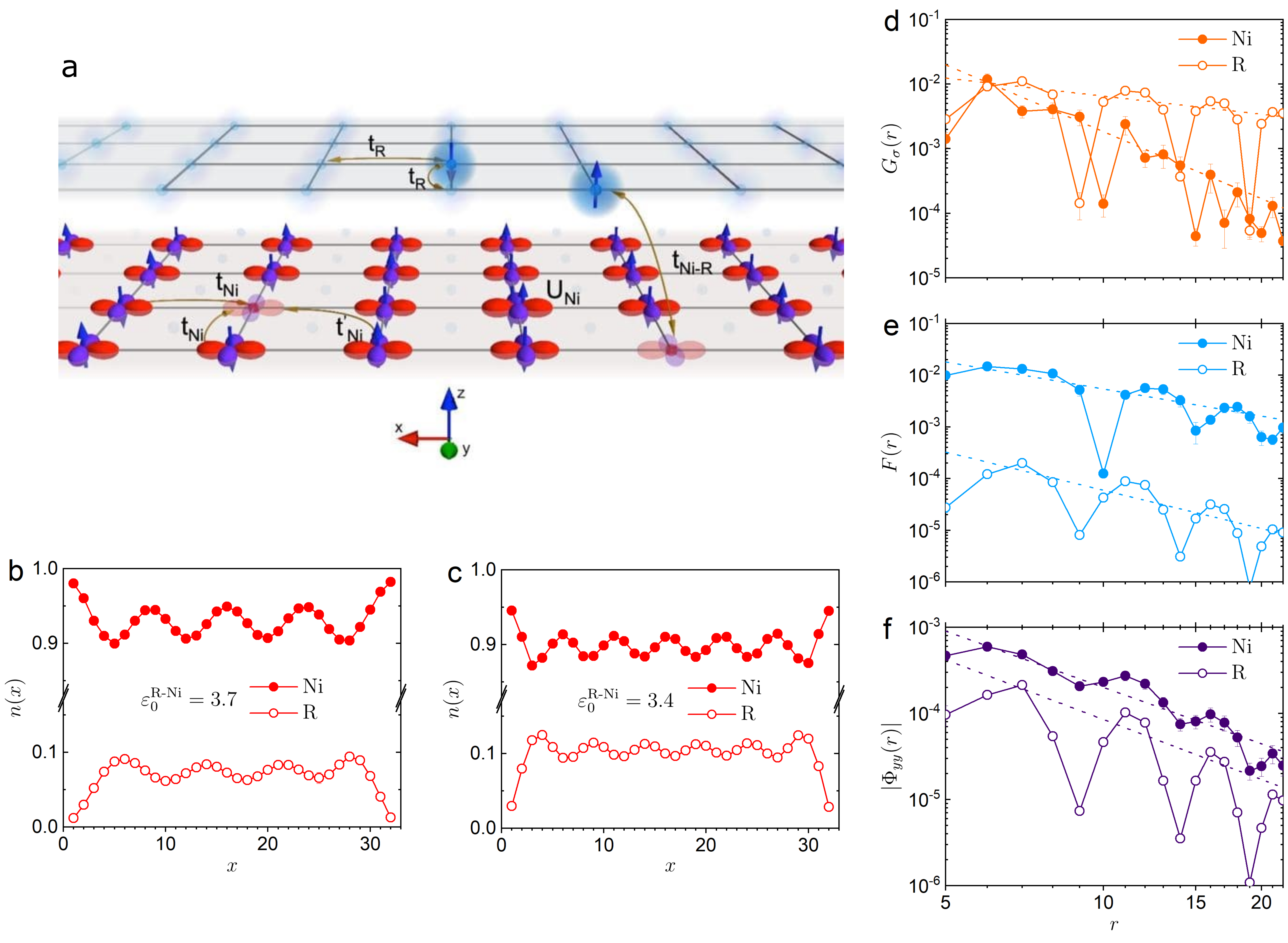}
    \caption{\textbf{a} Schematic two-band Hubbard model on a four-leg square 
    cylinder geometry, with periodic boundary condition along the short $y$-direction and open boundary condition along the longer $x$-direction. Here, $t_{Ni}$ and $t'_{Ni}$ represent the nearest-neighbour (NN) and next-nearest-neighbour (NNN) hopping within the Ni layer, respectively. $t_R$ is the NN hopping within the $R$-layer, and $t_{\text{Ni}-R}$ is the hopping between the two layers. \textbf{b-c} Charge density profile for $\varepsilon_0^{R\text{-Ni}}=3.7$ and $\varepsilon_0^{R\text{-Ni}}=3.4$ at half-filling. \textbf{d-f} Correlation functions measured on the Ni- and $R$-layers, respectively, for $\varepsilon_0^{R\text{-Ni}}=3.4$ at half-filling. The power-law fitting functions are represented by dashed lines.}
\label{Fig.HF3.4p3.7}
\end{figure*}

We consider the two-band Hubbard model proposed in Ref.\cite{Hepting2020}, with some simplifications to study the effects of strongly correlated electrons without overloading model parameters. The model that we use is depicted in Fig.\ref{Fig.HF3.4p3.7}a, whose Hamiltonian is defined as $H=H_{\text{Ni}}+H_{R}+H_{\text{Ni}-R}$. Specifically,
\begin{widetext}
\begin{eqnarray}
    &H_{\text{Ni}} = -t_{\text{Ni}}\sum_{\begin{subarray}{c}\langle ii'\rangle, \sigma
    \end{subarray}}(\hat{c}^{[\text{Ni}]\dagger}_{i\sigma}\hat{c}^{[\text{Ni}]}_{i'\sigma}+h.c.)-t'_{\text{Ni}}\sum_{\begin{subarray}{c} \langle\langle ii'\rangle\rangle,\sigma \end{subarray}}(\hat{c}^{[\text{Ni}]\dagger}_{i\sigma}\hat{c}^{[\text{Ni}]}_{i'\sigma}+h.c.)+U_{\text{Ni}}\sum_{i}\hat{n}^{[\text{Ni}]}_{i\uparrow}\hat{n}^{[\text{Ni}]}_{i\downarrow}, \\
    &H_{R} = -t_{R}\sum_{\begin{subarray}{c}\langle jj' \rangle, \sigma \end{subarray}}(\hat{c}^{[R]\dagger}_{j\sigma}\hat{c}^{[R]}_{j'\sigma}+h.c.)+\varepsilon_{0}^{R\text{-Ni}}\sum_{j}\hat{n}_{j}^{[R]},\\
    &H_{\text{Ni-}R} = -t_{\text{Ni-}R}\sum_{ij, \sigma}(\hat{c}^{[\text{Ni}]\dagger}_{i\sigma}\hat{c}^{[R]}_{j\sigma}+h.c.).
          \label{Hamiltonian}
\end{eqnarray}
\end{widetext}
Here, $\hat{c}^{[l]\dagger}_{i\sigma}$($\hat{c}^{[l]}_{i\sigma}$) is the electron creation (annihilation) operator with spin-$\sigma$ ($\sigma=\uparrow$, $\downarrow$) on site $i=(x,y)$ in the $l$-layer ($l $: Ni, $R$), and $\hat{n}^{[l]}_i$ is the electron number operator. $H_{\text{Ni}}$ denotes the Hamiltonian of the Ni $3d_{x^2-y^2}$ orbital. In the present study, we set $t_{\text{Ni}}=1$ as the energy unit and consider $t'_{\text{Ni}}=-0.25$ \cite{Hepting2020,Been2020}, where $\langle ii'\rangle$ and $\langle\langle ii'\rangle\rangle$ denote the nearest-neighbor (NN) and next-nearest-neighbor (NNN) sites in the Ni layer. The on-site Coulomb repulsion is set to $U_{\text{Ni}}=8$ in units of $t_{\text{Ni}}$. While this represents a typical value used for studies associated with the cuprates~\cite{Nomura2019prb}, it is in the intermediate coupling regime, between weak coupling and the Heisenberg limit, and consistent with constrained random phase approximation (cRPA) estimates~\cite{cRPAPRB2019,cRPAPRB2020}. $H_{R}$ is the Hamiltonian for the rare-earth $5d$ orbital ($R$: La, Nd, Pr etc.), and $H_{\text{Ni}-R}$ is the hybridization between Ni- and $R$-layers. We set $t_{R}=0.5$ and $t_{\text{Ni}-R}=0.07$, respectively, to mimic the actual band dispersion as in Ref.\cite{Hepting2020}.
$\varepsilon_{0}^{R\text{-Ni}}$ denotes the on-site energy difference between the $R$ and Ni bands. The on-site energy difference varies between different infinite-layer nickelate compounds, and also might change with chemical doping. Here, the value originates from Wannier downfolding the density functional theory (DFT) paramagnetic solution~\cite{Hepting2020}. Tuning the value of the on-site energy difference controls the size of the electron pocket, as well as the dominant order at half-filling. We take this on-site energy difference as a tunable parameter to adjust the relative electron density between the $R$ and Ni-bands and the $R$-pocket size compared to experiments.

\begin{table}[htbp]
\centering 
\begin{tabular}{p{1.2cm}<{\centering}  p{1.2cm}<{\centering}  p{1.2cm}<{\centering}  p{1.2cm}<{\centering}  p{1.2cm}<{\centering}  p{2.0cm}<{\centering} }
\hline\hline
$t_{\text{Ni}}$ & $t'_{\text{Ni}}$ & $U_{\text{Ni}}$ & $t_{\text{Ni-}R}$ & $t_{R}$ & $\varepsilon_0^{R\text{-Ni}}$\\ 
\hline 
 $1$ & $-0.25$ & $8$ & $0.07$ & $0.5$ & $3.2-4.0$ \\
\hline\hline
\end{tabular}
\caption{Summary of model parameters.}\label{Table:SummaryHamiltonian}
\end{table}

To reduce the computational complexity, we consider an inter-layer NN hopping in our model calculations. By symmetry, such an inter-layer NN hopping between Ni $d_{x^2-y^2}$ and $R$ $d_{z^2}$ should be zero; and a NNN hopping would be nonzero to leading order in the hybridization. However, since the electrons in the $R$-layer behave like a free Fermi gas due to their low electron density and high mobility, the details of the inter-layer hopping will not affect the leading role of the rare-earth $5d$ orbital. It is worth emphasizing that such a simplification will not change the ground state properties. This is confirmed further by the negligible difference between the charge density distribution computed separately from the $H_{\text{Ni}-R}$ that includes only NNN or NN hopping terms (details provided in the Supplementary Material~\cite{splm}). 

The lattice geometry used in our simulation is depicted in Fig.\ref{Fig.HF3.4p3.7}a, where $\hat{x}=(1,0)$ and $\hat{y}=(0,1)$ denote the two basis vectors of the square lattice. We consider a cylindrical geometry with periodic (open) boundary conditions in the $\hat{y}$ ($\hat{x}$) direction in each layer. We focus on four-leg cylinders with width $L_y=4$ and length up to $L_x=32$, where $L_y$ and $L_x$ are the number of sites along the $\hat{y}$ and $\hat{x}$ directions, respectively. The total number of sites of the system is $2N$ where $N=L_x\times L_y$. The hole doping concentration away from half-filling is defined as $\delta=N_h/N$, where $N_h$ is the number of doped holes ($N_h=0$ at half-filling). We keep up to $m=25,000$ states in our simulations with a typical truncation error $\epsilon\sim 10^{-6}$. The on-site energy difference $\varepsilon_0^{R\text{-Ni}}$ plays a vital role in controlling the size of the $R$-band electron pocket. For $\varepsilon_0^{R\text{-Ni}}$ in the range $3.4$ to $3.7$ the size of the rare-earth electron pocket matches well to experimental results~\cite{Li2020}. We focus primarily on $\varepsilon_0^{R\text{-Ni}}=3.4$ in this study to make a direct connection with experiment; however, we will show results for different values of $\varepsilon_0^{R\text{-Ni}}$ to aid our discussion. 
\section{Results}
\subsection{Half-filling}

The main results for the two-orbital Hubbard model at half-filling for the parent compound are shown in Fig. \ref{Fig.HF3.4p3.7}. Figures \ref{Fig.HF3.4p3.7}b and \ref{Fig.HF3.4p3.7}c show the charge density profile, defined as $n(x)=\sum_{y=1}^{L_y}\langle \hat{n}^{[l]}_i\rangle/L_y$, for the Ni- and $R$-layers, respectively. Here $\hat{n}^{[l]}_i$ is the electron number operator on site $i=(x,y)$ where $x$ and $y$ are the rung and row indices of the $l$-layer ($l=$ Ni, $R$). It is clear that a small fraction of electrons transfer from the Ni-layer to the $R$-layer, which gives rise to ``self-doped'' holes. The electron density in the $R$-layer is approximately $6.87\%$ for $\varepsilon_0^{R\text{-Ni}}=3.7$ and $10.04\%$ for $\varepsilon_0^{R\text{-Ni}}=3.4$, respectively. These densities are consistent with Hall coefficient measurements of the size of the rare-earth electron pocket $\sim 8\%$ per formula unit for NdNiO$_2$ \cite{Li2020}. Although the non-interacting $R$-band has much higher energy than the Ni band (see Supplementary Material~\cite{splm} for more details), the strong on-site Hubbard $U_{\text{Ni}}$ splits the Ni-band into upper and lower Hubbard bands, significantly broadening the interacting Ni bandwidth.

In Fig.~\ref{Fig.HF3.4p3.7}b-c, we see that a CDW has formed in both layers due to the modulation of the self-doped holes in the Ni-layer and the itinerant electrons in the $R$-layer. The CDW is consistent with nearly ``half-filed'' stripes in the Ni-layer, reminiscent of the charge stripes of the single-band Hubbard model with $t'$, although at finite doping~\cite{Jiang2020Hub,Jiang2018}. The spatial decay of the CDW correlations at long distances is dominated by a power-law with the Luttinger exponent $K_c$. This exponent can be obtained by fitting the charge density oscillations (Friedel oscillations)\cite{White2002,cdwosc2015prb} induced by the boundaries of the cylinder
\begin{equation}
n(x) \approx n_0+\frac{A\cos(Q_c x+\phi_{1})}{[L_{\text{eff}} \sin(\pi x/L_{\text{eff}}+\phi_{2})]^{K_c/2}} \label{eq.cdwfit}
\end{equation}
where $A$ is a non-universal amplitude, $\phi_{1}$ and $\phi_{2}$ are phase shifts, $n_0=1-\delta$ is the mean density, and $Q_c$ is the wavevector. An effective length of $L_{\text{eff}}\sim L_x$ best describes our results at half-filling; and the extracted exponent $K_c=1.35(5)$ for $\varepsilon_0^{R\text{-Ni}}=3.4$ in the Ni-layer.

Our results suggest that the overall ground state of the system for $\varepsilon_0^{R\text{-Ni}}=3.4$ and $3.7$ is consistent with a LL, as evidenced by the slow decay of the single-particle Green function in the $R$-layer, $G_\sigma(r)=|\langle \hat{c}^{[l]\dagger}_{i_0,\sigma} \hat{c}^{[l]}_{i_0+r,\sigma}\rangle|$, in contrast to the single-band Hubbard model at 1/2 filling. Here, $i_0=(x_0,y_0)$ is the reference site, $x_0\sim L_x/4$, and $r$ is the distance between two sites in the $x$-direction in each layer. For example, $G_\sigma(r)$ in the $R$-layer for $\varepsilon_0^{R\text{-Ni}}=3.4$ (shown in Fig.\ref{Fig.HF3.4p3.7}d) can be fit well by a power-law $G_\sigma(r)\sim r^{-K_G}$ with the Luttinger exponent $K_G= 0.9(2)$. This is the overall leading correlation which dominates over other correlations including the CDW correlation in the Ni-layer. On the contrary, $G_\sigma(r)$ in the Ni-layer decays much more rapidly, in fact consistent with an exponential decay $G_\sigma(r)\sim e^{-r/\xi_G}$ with a correlation length $\xi_G=3.5(3)$ in units of the lattice spacing. An alternative power-law fit $G_\sigma(r)\sim r^{-K_G}$ to the single-particle Green function in the Ni-layer yields $K_G=3.4(3)$, which is much larger than the expected exponent $K_G\sim 1$ for a LL in the Ni-layer.

For completeness, we have also calculated the spin-spin correlation function $F(r) = |\langle \mathbf{S}^{[l]}_{i_0}\cdot \mathbf{S}^{[l]}_{i_0+r}\rangle|$. As shown in Fig.\ref{Fig.HF3.4p3.7}e, $F(r)$ in the $R$-layer appears to decay as a power-law $F(r)\sim r^{-K_s}$, but with a large exponent $K_s=2.4(3)$, consistent with LL behavior. The spin-spin correlation function $F(r)$ in the Ni-layer is shown in Fig.\ref{Fig.HF3.4p3.7}e. While it is difficult to distinguish between power-law and exponential decay due to finite-size effects, we find that the behaviour of $F(r)$ of the lightly doped single-band Hubbard model on four-leg square cylinders, where $F(r)$ decays exponentially at long distances although dominates at short distances~\cite{Jiang1424}.

To test the possibility of superconductivity, we also calculate the equal-time SC pair-pair correlations defined as $\Phi_{\alpha\beta}(r)=\langle\Delta^{\dagger}_{\alpha}(x_0,y)\Delta_{\beta}(x_0+r,y)\rangle$, where $\Delta^{\dagger}_{\alpha}(i)=\frac{1}{\sqrt{2}}[\hat{c}^{[l]\dagger}_{i\uparrow}\hat{c}^{[l]\dagger}_{i+\alpha\downarrow}-\hat{c}^{[l]\dagger}_{i\downarrow}\hat{c}^{[l]\dagger}_{i+\alpha\uparrow}]$ is the spin-singlet pair-field creation operator. Here, $\alpha=x/y$ denotes the bond orientation and $r$ is the distance between two bonds along the $x$ direction. As expected for a LL, $\Phi_{\alpha\beta}(r)$ in the $R$-layer decays as a power-law, i.e., $|\Phi_{yy}|\sim r^{-K_{sc}}$ where $K_{sc}=2.3(3)$, as shown in Fig. \ref{Fig.HF3.4p3.7}f. $\Phi_{\alpha\beta}(r)$ in the Ni-layer shows similar behavior to $F(r)$; and it is difficult to precisely determine its decay at long distances due to finite-size effects. However, a power-law fit in the Ni-layer gives $K_{sc}=2.2(3)$, which suggests that the SC susceptibility is unlikely to diverge. Taken together, the ground state of the two-band Hubbard model in this range of on-site energy at half-filling is more consistent with a metallic LL, but is unlikely to be superconducting.

\subsection{CDW dependence on $\varepsilon_0^{R\text{-Ni}}$ and doping}

\begin{figure*}[tbh!]
\centering
    \includegraphics[width=0.8\linewidth]{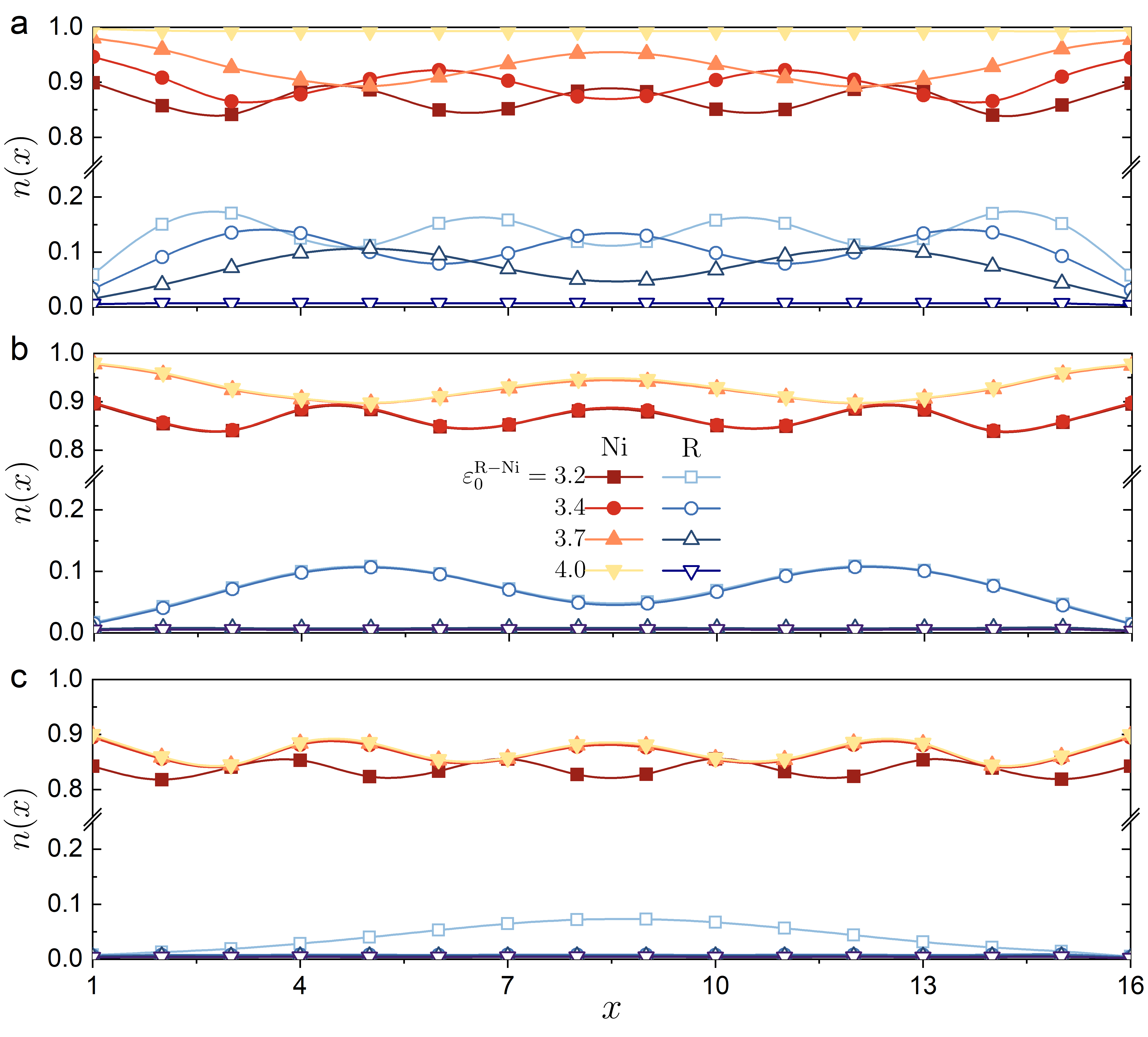}
    \caption{Charge density profile at \textbf{a} half-filling, \textbf{b} $\delta=6.25\%$, and \textbf{c} $12.5\%$ hole doping.}\label{Fig.CDW_Lx16}
\end{figure*}

Figure~\ref{Fig.CDW_Lx16} shows the DMRG results of the on-site energy difference $\varepsilon_0^{R\text{-Ni}}$ and doping dependence of the CDW periodicity in the Ni- and $R$-layers. At half-filling (Fig.~\ref{Fig.CDW_Lx16}a), smaller on-site energy differences between the two layers lead to shorter CDW wavelengths. The wavevectors for $\varepsilon_0^{R\text{-Ni}}=3.2$, $3.4$, and $3.7$ are $Q_c=\pi/2$, $3\pi/8$ and $\pi/4$, respectively. For larger energy differences, {\it e.g.}~$\varepsilon_0^{R\text{-Ni}}=4.0$, there are no clearly discernible charge modulations. The recent observations of charge order in the infinite-layer nickelates show an incommensurate wavevector of $Q_c\sim 2\pi/3$ with a wavelength $\sim 3$ lattice spacings\cite{Matteo2021arxiv}, which could be reproduced with a slightly smaller value of $\varepsilon_0^{R\text{-Ni}} \sim 3.0$. Importantly, the CDW modulations at half-filling in the two layers lock together at the same wavevector with an anti-phase pattern between the Ni and the $R$ charge modulation, when the self-doped hole concentration in the Ni-layer equals the charge filling concentration in the $R$-layer. This locked CDW modulations with the same wavevector between Ni- and $R$-layers were also observed in RIXS measurement for the infinite-layer LaNiO${_2}$.\cite{Matteo2021arxiv}

With light doping, this locking of the CDW between the two layers is released, as shown in Fig.\ref{Fig.CDW_Lx16}b for $\delta=6.25\%$, where the periodicity of the CDW becomes different in the two layers for the same $\varepsilon_0^{R\text{-Ni}}$. Additional doping introduces more holes in the Ni-layer and depletes charge carriers from the $R$-layer, {\it e.g.} $\delta=12.5\%$, as shown in Fig.\ref{Fig.CDW_Lx16}c. At this doping concentration, the Ni-layer shows more clear charge stripes, while the electron density in the $R$-layer is very low but nonzero. It remains an open question whether the CDW, as well as superconducting correlations, would become the leading order for some special value of the on-site energy difference $\varepsilon_0^{R\text{-Ni}}$ even at half-filling. 

\subsection{Finite hole doping}

\begin{figure*}[tb] 
\centering
    \includegraphics[width=1\linewidth]{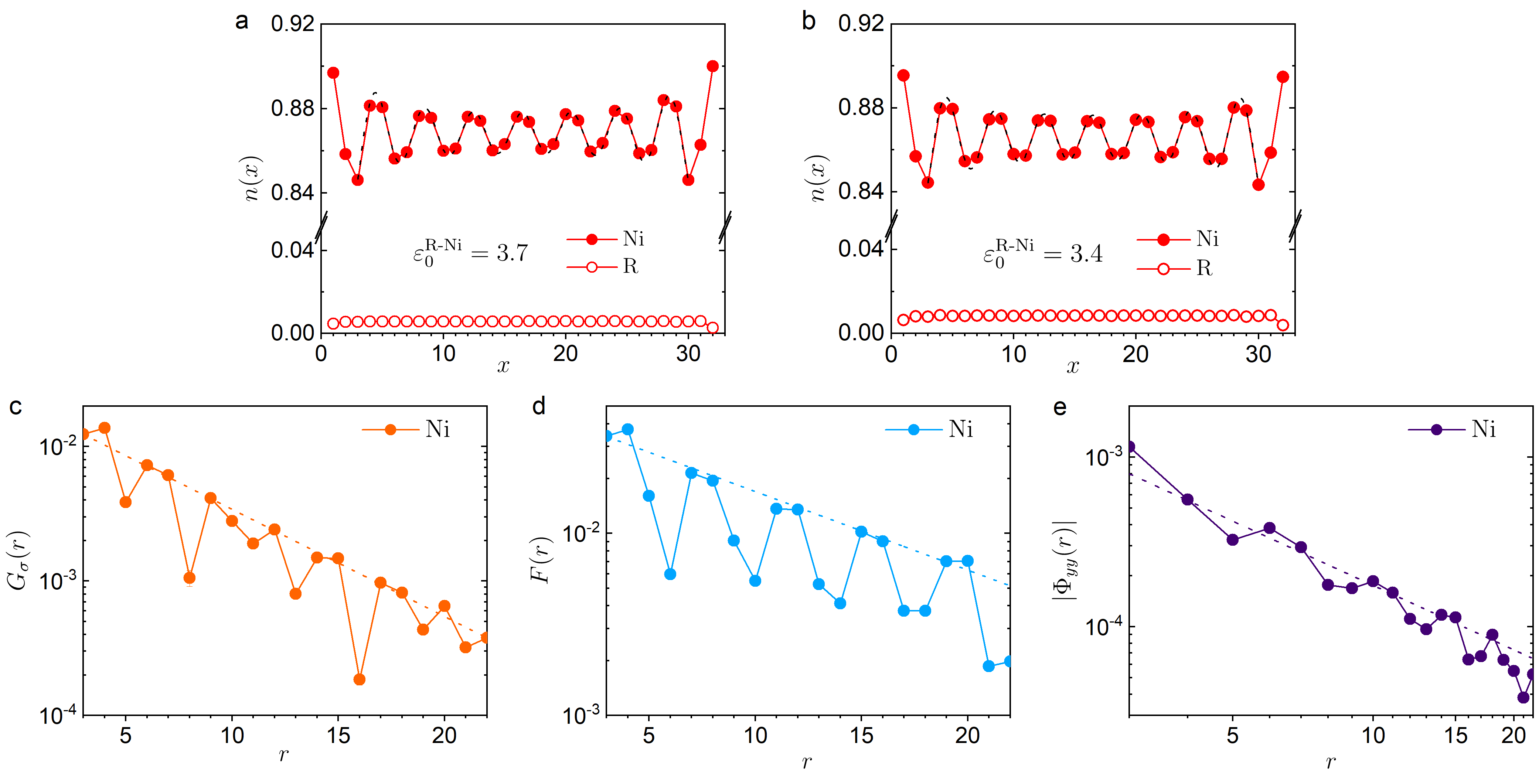}
    \caption{\textbf{a-b} The charge density profile for $\varepsilon_0^{R\text{-Ni}}=3.7$ and $\varepsilon_0^{R\text{-Ni}}=3.4$ at $\delta=12.5\%$ doping. Fits to Eq.~\ref{eq.cdwfit} are represented by dashed lines. \textbf{c-e} Correlation functions measured in the Ni-layer for $\varepsilon_0^{R\text{-Ni}}=3.4$ at $\delta=12.5\%$ doping. The power-law fitting functions are represented by dashed lines.}
\label{Fig.IR8FD3.4p3.7}
\end{figure*}

\begin{table*}[htbp]
\centering 
\begin{tabular}{ p{2.5cm}<{\centering}  p{2.5cm}<{\centering}  p{2.5cm}<{\centering}  p{2.5cm}<{\centering}  p{2.5cm}<{\centering}  p{2.5cm}<{\centering} }
\hline\hline
 $\varepsilon_0^{R\text{-Ni}}$  &  $K_c$ & $K_{sc}$ & $\xi_s$ & $\xi_G$ & $c$\\ 
\hline 
  3.7 & 1.0(2)  & 1.1(1) & 10.8(2) & 4.8(2) & 1.2(3)\\
  3.4 & 0.9(1) & 1.3(2) & 10.0(6) & 5.8(1) &  1.3(2)\\
\hline\hline
\end{tabular}
\caption{This table lists the Luttinger exponents $K_c$ and $K_{sc}$, the correlation lengths (in units of lattice spacing) $\xi_s$ and $\xi_{G}$, as well as the central charge $c$ at a doping concentration $\delta=12.5\%$. The system length is fixed to $L_x=32$. Those exponents (correlation lengths) are extracted from the algebraically (exponentially) decaying correlation functions measured on the Ni-layer.}\label{Table:SummaryFDIR8}
\end{table*}

Experiments have shown that the phase diagram of infinite-layer Nd$_{1-x}$Sr$_x$NiO$_2$ possess a SC dome in the range of $12\% < \delta < 25\%$ hole doping~\cite{Li2020,Zeng2020}. In our search for superconductivity in the two-band model, we focus on the underdoped region with $\delta=12.5\%$. The main results are shown in Fig.~\ref{Fig.IR8FD3.4p3.7}. Nearly all of the charge carriers reside in the Ni-layer for both $\varepsilon_0^{R\text{-Ni}}=3.7$ and $3.4$, consistent with experiments~\cite{Rossi2020, Li2020}. It may be worth mentioning again that the residual electron density remains nonzero in the $R$-layer, {\it e.g.}~around $0.01$ for $\varepsilon_0^{R\text{-Ni}}=3.4$. This would appear to agree with the discovery of a non-negligible contribution from the Nd $5d$ electron pocket, reported from RIXS measurements at the Ni $L$-edge for Sr-doped nickelates~\cite{Rossi2020}. At $\delta=12.5\%$, the results for the two-band model agree reasonably well with previous studies of the single-band Hubbard~\cite{Jiang1424,Jiang2020Hub,Kitatani2020,Jiang2018} and $t$-$J$ models~\cite{Jiang2018}, where the ground state of the system possess ``half-filled" charge stripes (see Fig.~\ref{Fig.IR8FD3.4p3.7}a-b) with an ordering vector $Q_c=4\pi\delta$ of wavelength $\lambda_c=1/2\delta$ -- half of a doped hole per CDW unit cell.

Distinct from half-filling, the single-particle Green functions $G_\sigma(r)$ at $\delta=12.5\%$ in both the Ni- and $R$-layers are short-ranged, decaying exponentially as $G_\sigma(r)\propto e^{-r/\xi_G}$, shown in Fig.~\ref{Fig.IR8FD3.4p3.7}c. We provide the value of $\xi_G$ in Table \ref{Table:SummaryFDIR8}. To measure the magnetic properties of the system, we also calculated the spin-spin correlations $F(r)$, shown in Fig.~\ref{Fig.IR8FD3.4p3.7}d. It is clear that $F(r)$ also decays exponentially as $F(r)\propto e^{-r/\xi_s}$, with an extracted correlation length $\xi_{s}\approx 10$ lattice spacings. This is similar to the value in previous studies of the single-band Hubbard model on the square lattice~\cite{Jiang1424}.

Unlike both $G_\sigma(r)$ and $F(r)$, we find that the equal-time SC and CDW correlations are quasi-long-ranged. The spatial decay of the SC correlation at long distances is dominated by a slow power-law decay $\Phi_{\alpha\beta}(r)\sim r^{-K_{sc}}$ with a Luttinger exponent $K_{sc}\sim 1$. Results are shown in Fig.~\ref{Fig.IR8FD3.4p3.7}d and Table \ref{Table:SummaryFDIR8}. The pairing symmetry of the SC correlations is consistent with plaquette $d$-wave on four-leg square cylinders, which is evidenced by the fact that the dominant SC correlation $\Phi_{yy}(r)$ changes sign around the cylinder~\cite{Dodaro2017,Jiang1424,Jiang2018,Jiang2020Hub,Chung2020}. The spatial decay of the CDW correlations at long distances also is dominated by a power-law, with corresponding Luttinger exponent $K_c$. The exponent $K_c$ can be obtained by fitting the charge density oscillations (Friedel oscillations). Note that an effective length $L_{\text{eff}}\sim L_x-2$ best describes our results at $12.5\%$ hole doping. The extracted exponent $K_c$, as summarized in Table \ref{Table:SummaryFDIR8}, is consistent with $K_{c}\sim 1$. Taken together, our results suggest that the ground state of the system at $\delta=12.5\%$ is consistent with a Luther-Emery (LE) liquid~\cite{LE1974}, having the expected dual relationship $K_c\sim 1/K_{sc}$ within numerical uncertainty.

As a further test for the LE liquid, the system should have a single gapless charge mode with central charge $c=1$. To demonstrate this, we have calculated the von Neumann entanglement entropy $S(x)=-\mathrm{Tr} {\rho_x \ln \rho_x}$, where $\rho_x$ is the reduced density matrix of a subsystem with length $x$. The extracted central charge $c$ (see Table \ref{Table:SummaryFDIR8}) is reasonably close to $c=1$ (see Supplementary Material~\cite{splm} for more details). Therefore, our results for the two-band Hubbard model on four-leg square cylinders at $\delta=12.5\%$ show that the ground state is a LE liquid with a strongly diverging SC susceptibility $\chi_{sc}\sim T^{-(2-K_{sc})}$ as the temperature $T\rightarrow 0$.

\section{Discussion}%
Our DMRG results suggest that the itinerant electrons in the rare-earth $5d$ orbital play a vital role in leading the ground state of the two-band Hubbard model to behave distinctly from the single-band Hubbard model at half-filling. Surprisingly, we have seen nearly ``half-filled" charge stripes in the Ni layer even at half-filling. It remains possible that manipulating the model parameters, such as the on-site energy difference, may enhance the SC pair-pair correlation function in the Ni layer, leading alternatively to a LE liquid at half-filling. For $\delta=12.5\%$ hole doping, the results for the two-band Hubbard model agree with previous studies of the single-band Hubbard and $t$-$J$ models, as the rare-earth $5d$ band becomes almost empty. This interplay between the itinerant rare-earth $5d$ electrons and the Ni $3d_{x^2-y^2}$ orbital coincides with the experimental observation that the $5d$ electron pocket diminishes and the $3d$ hole pocket becomes dominant when the doping level increases to approximately $12\%$, also where the SC dome starts to emerge~\cite{Li2019,Li2020,Zeng2020,Osada2020}. Our DMRG results at $\delta=12.5\%$ may also help to explain the similarity of the phase diagram between the infinite-layer nickelates and cuprates in the overdoped regime and why the SC dome of infinite-layer nickelates ends at $\delta \approx 25\%$, a typical value for cuprates.

Needless to say, there is a speculative leap from our results on quasi-1D cylinders to the 2D limit. Even for the single-band Hubbard model where accurate ground state properties, such as superconductivity, are still under debate on cylinders wider than four-leg for negative $t'$. However, we feel that the present results may be a precursor of the solution of the corresponding 2D problem, given that the Luttinger (Luther-Emery) liquid is the 1D descendent of the Fermi liquid (superconducting phase) in two or higher dimensions. The model on narrow cylinders like four-leg shows consistency with experimental observations, especially charge order in the undoped parent materials. It would be interesting in the future to study how the system evolves if expanded to the 2D and 3D limits. Here, we have focused on half-filling and under-doping. It also will be interesting to study the on-site energy difference and higher doping, including optimal doping. Meanwhile, we expect that more unbiased numerical simulations of this model could be performed in the future to better understand its physical properties at both zero and finite temperatures and the dynamical spin and charge responses. Answering these questions could shed new light on the mechanism of high-temperature superconductivity in both nickelates and cuprates.

\textbf{Acknowledgments:} We would like to thank Steven Kivelson, Bai-Yang Wang, Harold Hwang and Srinivas Raghu for insightful discussions. This work was supported by the Department of Energy, Office of Science, Basic Energy Sciences, Materials Sciences and Engineering Division, under Contract No. DE-AC02-76SF00515. Parts of the computing for this project were performed on the National Energy Research Scientific Computing Center (NERSC), a US Department of Energy Office of Science User Facility operated under Contract No. DE-AC02-05CH11231. Parts of the computing for this project were performed on the Sherlock cluster.

\textbf{Author contributions:} C.J.J. and C.P. conceived the study. C.P. and H.C.J. performed numerical simulations. C.P. analyzed data under the supervision of H.C.J. and C.J.J. All authors assisted in data interpretation and contributed to the writing of the manuscript.

\textbf{Competing interests:} The authors declare no competing interests.

\textbf{Data and materials availability:} The main data supporting the findings of this study are available within the main text and the supplementary materials. Extra data are available from the corresponding author upon reasonable request. The codes implementing the calculations of this study are available from the corresponding author upon request.


\begin{thebibliography}{10}

\bibitem{Li2019}
Danfeng Li, Kyuho Lee, Bai~Yang Wang, Motoki Osada, Samuel Crossley, Hye~Ryoung
  Lee, Yi~Cui, Yasuyuki Hikita, and Harold~Y. Hwang.
\newblock Superconductivity in an infinite-layer nickelate.
\newblock {\em Nature}, 572:624, Aug 2019.

\bibitem{Osada2020}
Motoki Osada, Bai~Yang Wang, Kyuho Lee, Danfeng Li, and Harold~Y. Hwang.
\newblock {Phase diagram of infinite layer praseodymium nickelate
  ${\mathrm{Pr}}_{1\ensuremath{-}x}{\mathrm{Sr}}_{x}{\mathrm{NiO}}_{2}$ thin
  films}.
\newblock {\em Phys. Rev. Materials}, 4:121801, Dec 2020.

\bibitem{Zeng2020}
Shengwei Zeng, Chi~Sin Tang, Xinmao Yin, Changjian Li, Mengsha Li, Zhen Huang,
  Junxiong Hu, Wei Liu, Ganesh~Ji Omar, Hariom Jani, Zhi~Shiuh Lim, Kun Han,
  Dongyang Wan, Ping Yang, Stephen~John Pennycook, Andrew T.~S. Wee, and
  Ariando Ariando.
\newblock {Phase Diagram and Superconducting Dome of Infinite-Layer
  ${\mathrm{Nd}}_{1\ensuremath{-}x}{\mathrm{Sr}}_{x}{\mathrm{NiO}}_{2}$ Thin
  Films}.
\newblock {\em Phys. Rev. Lett.}, 125:147003, Oct 2020.

\bibitem{Li2020}
Danfeng Li, Bai~Yang Wang, Kyuho Lee, Shannon~P. Harvey, Motoki Osada, Berit~H.
  Goodge, Lena~F. Kourkoutis, and Harold~Y. Hwang.
\newblock {Superconducting Dome in
  ${\mathrm{Nd}}_{1\ensuremath{-}x}{\mathrm{Sr}}_{x}{\mathrm{NiO}}_{2}$
  Infinite Layer Films}.
\newblock {\em Phys. Rev. Lett.}, 125:027001, Jul 2020.

\bibitem{Osada2021}
Motoki Osada, Bai~Yang Wang, Berit~H. Goodge, Shannon~P. Harvey, Kyuho Lee,
  Danfeng Li, Lena~F. Kourkoutis, and Harold~Y. Hwang.
\newblock Nickelate superconductivity without rare-earth magnetism:
  (la,sr)nio2.
\newblock {\em Advanced Materials}, n/a(n/a):2104083.

\bibitem{2022arXiv220302580L}
Kyuho {Lee}, Bai~Yang {Wang}, Motoki {Osada}, Berit~H. {Goodge}, Tiffany~C.
  {Wang}, Yonghun {Lee}, Shannon {Harvey}, Woo~Jin {Kim}, Yijun {Yu}, Chaitanya
  {Murthy}, Srinivas {Raghu}, Lena~F. {Kourkoutis}, and Harold~Y. {Hwang}.
\newblock {Character of the ``normal state'' of the nickelate superconductors}.
\newblock {\em arXiv e-prints}, page arXiv:2203.02580, March 2022.

\bibitem{Rossi2020}
M.~Rossi, H.~Lu, A.~Nag, D.~Li, M.~Osada, K.~Lee, B.~Y. Wang, S.~Agrestini,
  M.~Garcia-Fernandez, J.~J. Kas, Y.-D. Chuang, Z.~X. Shen, H.~Y. Hwang,
  B.~Moritz, Ke-Jin Zhou, T.~P. Devereaux, and W.~S. Lee.
\newblock Orbital and spin character of doped carriers in infinite-layer
  nickelates.
\newblock {\em Phys. Rev. B}, 104:L220505, Dec 2021.

\bibitem{Goodgee2021pnas}
Berit~H. Goodge, Danfeng Li, Kyuho Lee, Motoki Osada, Bai~Yang Wang, George~A.
  Sawatzky, Harold~Y. Hwang, and Lena~F. Kourkoutis.
\newblock {Doping evolution of the Mott{\textendash}Hubbard landscape in
  infinite-layer nickelates}.
\newblock {\em Proceedings of the National Academy of Sciences}, 118(2), 2021.

\bibitem{Hepting2020}
M.~Hepting, D.~Li, and C.J. et~al. Jia.
\newblock Electronic structure of the parent compound of superconducting
  infinite-layer nickelates.
\newblock {\em Nature Materials}, 19:381–385, Jan 2020.

\bibitem{Matteo2021arxiv}
Matteo {Rossi}, Motoki {Osada}, Jaewon {Choi}, Stefano {Agrestini}, Daniel
  {Jost}, Yonghun {Lee}, Haiyu {Lu}, Bai~Yang {Wang}, Kyuho {Lee}, Abhishek
  {Nag}, Yi-De {Chuang}, Cheng-Tai {Kuo}, Sang-Jun {Lee}, Brian {Moritz},
  Thomas~P. {Devereaux}, Zhi-Xun {Shen}, Jun-Sik {Lee}, Ke-Jin {Zhou},
  Harold~Y. {Hwang}, and Wei-Sheng {Lee}.
\newblock {A Broken Translational Symmetry State in an Infinite-Layer
  Nickelate}.
\newblock {\em Nat. Phys.}, 2022.

\bibitem{ZhouKeJin2021arxiv}
Charles~C. {Tam}, Jaewon {Choi}, Xiang {Ding}, Stefano {Agrestini}, Abhishek
  {Nag}, Bing {Huang}, Huiqian {Luo}, Mirian {Garc{\'\i}a-Fern{\'a}ndez}, Liang
  {Qiao}, and Ke-Jin {Zhou}.
\newblock {Charge density waves in infinite-layer NdNiO$_2$ nickelates}.
\newblock {\em arXiv e-prints}, page arXiv:2112.04440, December 2021.

\bibitem{Krieger2021arxiv}
G.~{Krieger}, L.~{Martinelli}, S.~{Zeng}, L.~E. {Chow}, K.~{Kummer},
  R.~{Arpaia}, M.~{Moretti Sala}, N.~B. {Brookes}, A.~{Ariando}, N.~{Viart},
  M.~{Salluzzo}, G.~{Ghiringhelli}, and D.~{Preziosi}.
\newblock {Charge and spin order dichotomy in NdNiO$_2$ driven by SrTiO$_3$
  capping layer}.
\newblock {\em arXiv e-prints}, page arXiv:2112.03341, December 2021.

\bibitem{Zaanen1985prl}
J.~Zaanen, G.~A. Sawatzky, and J.~W. Allen.
\newblock Band gaps and electronic structure of transition-metal compounds.
\newblock {\em Phys. Rev. Lett.}, 55:418--421, Jul 1985.

\bibitem{Chen1991prl}
C.~T. Chen, F.~Sette, Y.~Ma, M.~S. Hybertsen, E.~B. Stechel, W.~M.~C. Foulkes,
  M.~Schulter, S-W. Cheong, A.~S. Cooper, L.~W. Rupp, B.~Batlogg, Y.~L. Soo,
  Z.~H. Ming, A.~Krol, and Y.~H. Kao.
\newblock {Electronic states in
  ${\mathrm{La}}_{2\mathrm{\ensuremath{-}}\mathit{x}}$${\mathrm{Sr}}_{\mathit{x}}$${\mathrm{CuO}}_{4+\mathrm{\ensuremath{\delta}}}$
  probed by soft-x-ray absorption}.
\newblock {\em Phys. Rev. Lett.}, 66:104--107, Jan 1991.

\bibitem{Anisimov1999}
V.~I. Anisimov, D.~Bukhvalov, and T.~M. Rice.
\newblock Electronic structure of possible nickelate analogs to the cuprates.
\newblock {\em Phys. Rev. B}, 59:7901--7906, Mar 1999.

\bibitem{Lee2004}
K.-W. Lee and W.~E. Pickett.
\newblock {Infinite-layer $\mathrm{La}\mathrm{Ni}{\mathrm{O}}_{2}$:
  ${\mathrm{Ni}}^{1+}$ is not ${\mathrm{Cu}}^{2+}$}.
\newblock {\em Phys. Rev. B}, 70:165109, Oct 2004.

\bibitem{Hirayama2020}
Motoaki Hirayama, Terumasa Tadano, Yusuke Nomura, and Ryotaro Arita.
\newblock {Materials design of dynamically stable ${d}^{9}$ layered
  nickelates}.
\newblock {\em Phys. Rev. B}, 101:075107, Feb 2020.

\bibitem{Botana2020prx}
A.~S. Botana and M.~R. Norman.
\newblock {Similarities and Differences between ${\mathrm{LaNiO}}_{2}$ and
  ${\mathrm{CaCuO}}_{2}$ and Implications for Superconductivity}.
\newblock {\em Phys. Rev. X}, 10:011024, Feb 2020.

\bibitem{Wu2020prb}
Xianxin Wu, Domenico Di~Sante, Tilman Schwemmer, Werner Hanke, Harold~Y. Hwang,
  Srinivas Raghu, and Ronny Thomale.
\newblock {Robust ${d}_{{x}^{2}\ensuremath{-}{y}^{2}}$-wave superconductivity
  of infinite-layer nickelates}.
\newblock {\em Phys. Rev. B}, 101:060504, Feb 2020.

\bibitem{Meijw2020}
Hu~Zhang, Lipeng Jin, Shanmin Wang, Bin Xi, Xingqiang Shi, Fei Ye, and Jia-Wei
  Mei.
\newblock {Effective Hamiltonian for nickelate oxides
  ${\mathrm{Nd}}_{1\ensuremath{-}x}{\mathrm{Sr}}_{x}{\mathrm{NiO}}_{2}$}.
\newblock {\em Phys. Rev. Research}, 2:013214, Feb 2020.

\bibitem{Sakakibara2020prl}
Hirofumi Sakakibara, Hidetomo Usui, Katsuhiro Suzuki, Takao Kotani, Hideo Aoki,
  and Kazuhiko Kuroki.
\newblock {Model Construction and a Possibility of Cupratelike Pairing in a New
  ${d}^{9}$ Nickelate Superconductor
  $(\mathrm{Nd},\mathrm{Sr}){\mathrm{NiO}}_{2}$}.
\newblock {\em Phys. Rev. Lett.}, 125:077003, Aug 2020.

\bibitem{Lechermann2020}
Frank Lechermann.
\newblock Late transition metal oxides with infinite-layer structure:
  Nickelates versus cuprates.
\newblock {\em Phys. Rev. B}, 101:081110, Feb 2020.

\bibitem{WuX2020}
Xianxin Wu, Domenico Di~Sante, Tilman Schwemmer, Werner Hanke, Harold~Y. Hwang,
  Srinivas Raghu, and Ronny Thomale.
\newblock {Robust ${d}_{{x}^{2}\ensuremath{-}{y}^{2}}$-wave superconductivity
  of infinite-layer nickelates}.
\newblock {\em Phys. Rev. B}, 101:060504, Feb 2020.

\bibitem{Nomura2020}
Yusuke Nomura, Motoaki Hirayama, Terumasa Tadano, Yoshihide Yoshimoto, Kazuma
  Nakamura, and Ryotaro Arita.
\newblock {Formation of a two-dimensional single-component correlated electron
  system and band engineering in the nickelate superconductor
  ${\mathrm{NdNiO}}_{2}$}.
\newblock {\em Phys. Rev. B}, 100:205138, Nov 2019.

\bibitem{Jiang2020}
Mi~Jiang, Mona Berciu, and George~A. Sawatzky.
\newblock {Critical Nature of the Ni Spin State in Doped
  ${\mathrm{NdNiO}}_{2}$}.
\newblock {\em Phys. Rev. Lett.}, 124:207004, May 2020.

\bibitem{Wang2020prb}
Y.~Wang, C.-J. Kang, H.~Miao, and G.~Kotliar.
\newblock {Hund's metal physics: From ${\mathrm{SrNiO}}_{2}$ to
  ${\mathrm{LaNiO}}_{2}$}.
\newblock {\em Phys. Rev. B}, 102:161118, Oct 2020.

\bibitem{Werner2020prb}
Philipp Werner and Shintaro Hoshino.
\newblock Nickelate superconductors: Multiorbital nature and spin freezing.
\newblock {\em Phys. Rev. B}, 101:041104, Jan 2020.

\bibitem{Bhattacharyya2020prb}
Shinibali Bhattacharyya, P.~J. Hirschfeld, Thomas~A. Maier, and Douglas~J.
  Scalapino.
\newblock Effects of momentum-dependent quasiparticle renormalization on the
  gap structure of iron-based superconductors.
\newblock {\em Phys. Rev. B}, 101:174509, May 2020.

\bibitem{Liu2021prb}
Zhao Liu, Chenchao Xu, Chao Cao, W.~Zhu, Z.~F. Wang, and Jinlong Yang.
\newblock {Doping dependence of electronic structure of infinite-layer
  ${\mathrm{NdNiO}}_{2}$}.
\newblock {\em Phys. Rev. B}, 103:045103, Jan 2021.

\bibitem{Bjornson2021PRB}
Kristofer Bj\"ornson, Andreas Kreisel, Astrid~T. R\o{}mer, and Brian~M.
  Andersen.
\newblock Orbital-dependent self-energy effects and consequences for the
  superconducting gap structure in multiorbital correlated electron systems.
\newblock {\em Phys. Rev. B}, 103:024508, Jan 2021.

\bibitem{Zhang2020prr}
Hu~Zhang, Lipeng Jin, Shanmin Wang, Bin Xi, Xingqiang Shi, Fei Ye, and Jia-Wei
  Mei.
\newblock {Effective Hamiltonian for nickelate oxides
  ${\mathrm{Nd}}_{1\ensuremath{-}x}{\mathrm{Sr}}_{x}{\mathrm{NiO}}_{2}$}.
\newblock {\em Phys. Rev. Research}, 2:013214, Feb 2020.

\bibitem{Fu2019arxiv}
Ying {Fu}, Le~{Wang}, Hu~{Cheng}, Shenghai {Pei}, Xuefeng {Zhou}, Jian {Chen},
  Shaoheng {Wang}, Ran {Zhao}, Wenrui {Jiang}, Cai {Liu}, Mingyuan {Huang},
  XinWei {Wang}, Yusheng {Zhao}, Dapeng {Yu}, Fei {Ye}, Shanmin {Wang}, and
  Jia-Wei {Mei}.
\newblock {{Core-level x-ray photoemission and Raman spectroscopy studies on
  electronic structures in Mott-Hubbard type nickelate oxide NdNiO$_2$}}.
\newblock {\em arXiv e-prints}, page arXiv:1911.03177, November 2019.

\bibitem{Nomura2019prb}
Yusuke Nomura, Motoaki Hirayama, Terumasa Tadano, Yoshihide Yoshimoto, Kazuma
  Nakamura, and Ryotaro Arita.
\newblock {Formation of a two-dimensional single-component correlated electron
  system and band engineering in the nickelate superconductor
  ${\mathrm{NdNiO}}_{2}$}.
\newblock {\em Phys. Rev. B}, 100:205138, Nov 2019.

\bibitem{White1992}
Steven~R. White.
\newblock Density matrix formulation for quantum renormalization groups.
\newblock {\em Phys. Rev. Lett.}, 69:2863--2866, Nov 1992.

\bibitem{LE1974}
A.~Luther and V.~J. Emery.
\newblock Backward scattering in the one-dimensional electron gas.
\newblock {\em Phys. Rev. Lett.}, 33:589--592, Sep 1974.

\bibitem{Been2020}
Emily Been, Wei-Sheng Lee, Harold~Y. Hwang, Yi~Cui, Jan Zaanen, Thomas
  Devereaux, Brian Moritz, and Chunjing Jia.
\newblock Electronic structure trends across the rare-earth series in
  superconducting infinite-layer nickelates.
\newblock {\em Phys. Rev. X}, 11:011050, Mar 2021.

\bibitem{cRPAPRB2019}
Yusuke Nomura, Motoaki Hirayama, Terumasa Tadano, Yoshihide Yoshimoto, Kazuma
  Nakamura, and Ryotaro Arita.
\newblock Formation of a two-dimensional single-component correlated electron
  system and band engineering in the nickelate superconductor
  ${\mathrm{ndnio}}_{2}$.
\newblock {\em Phys. Rev. B}, 100:205138, Nov 2019.

\bibitem{cRPAPRB2020}
Motoaki Hirayama, Terumasa Tadano, Yusuke Nomura, and Ryotaro Arita.
\newblock Materials design of dynamically stable ${d}^{9}$ layered nickelates.
\newblock {\em Phys. Rev. B}, 101:075107, Feb 2020.

\bibitem{splm}
See the supplemental material at [url] for more numerical results and technical
  details.

\bibitem{Jiang2020Hub}
Yi-Fan Jiang, Jan Zaanen, Thomas~P. Devereaux, and Hong-Chen Jiang.
\newblock {Ground state phase diagram of the doped Hubbard model on the
  four-leg cylinder}.
\newblock {\em Phys. Rev. Research}, 2:033073, Jul 2020.

\bibitem{Jiang2018}
Hong-Chen Jiang, Zheng-Yu Weng, and Steven~A. Kivelson.
\newblock {Superconductivity in the doped $\mathit{t}\ensuremath{-}\mathit{J}$
  model: Results for four-leg cylinders}.
\newblock {\em Phys. Rev. B}, 98:140505, Oct 2018.

\bibitem{White2002}
Steven~R. White, Ian Affleck, and Douglas~J. Scalapino.
\newblock Friedel oscillations and charge density waves in chains and ladders.
\newblock {\em Phys. Rev. B}, 65:165122, Apr 2002.

\bibitem{cdwosc2015prb}
Michele Dolfi, Bela Bauer, Sebastian Keller, and Matthias Troyer.
\newblock Pair correlations in doped hubbard ladders.
\newblock {\em Phys. Rev. B}, 92:195139, Nov 2015.

\bibitem{Jiang1424}
Hong-Chen Jiang and Thomas~P. Devereaux.
\newblock {Superconductivity in the doped Hubbard model and its interplay with
  next-nearest hopping $t'$}.
\newblock {\em Science}, 365(6460):1424--1428, 2019.

\bibitem{Kitatani2020}
Motoharu Kitatani, Liang Si, Oleg Janson, Ryotaro Arita, Zhicheng Zhong, and
  Karsten Held.
\newblock {Nickelate superconductors — a renaissance of the one-band Hubbard
  model}.
\newblock {\em npj Quantum Materials}, 5:59, Aug 2020.

\bibitem{Dodaro2017}
John~F. Dodaro, Hong-Chen Jiang, and Steven~A. Kivelson.
\newblock {Intertwined order in a frustrated four-leg $t\ensuremath{-}J$
  cylinder}.
\newblock {\em Phys. Rev. B}, 95:155116, Apr 2017.

\bibitem{Chung2020}
Chia-Min Chung, Mingpu Qin, Shiwei Zhang, Ulrich Schollw\"ock, and Steven~R.
  White.
\newblock {Plaquette versus ordinary $d$-wave pairing in the ${t'}$-Hubbard
  model on a width-4 cylinder}.
\newblock {\em Phys. Rev. B}, 102:041106, Jul 2020.

\end{thebibliography}

\clearpage

\end{document}